# Self-assembly of iron nanoclusters on the $Fe_3O_4(111)$ superstructured surface

N. Berdunov, G. Mariotto, S. Murphy, I.V. Shvets
SFI Laboratories, Physics Dept., Trinity College, Dublin, Ireland

## Abstract

We report on the self-organized growth of a regular array of Fe nanoclusters on a nanopatterned magnetite surface. Under oxidizing preparation conditions the (111) surface of magnetite exhibits a regular superstructure with three-fold symmetry and a 42 Å periodicity. This superstructure represents an oxygen terminated (111) surface, which is reconstructed to form a periodically strained surface. This strain patterned surface has been used as a template for the growth of an ultrathin metal film. A Fe film of 0.5 Å thickness was deposited on the substrate at room temperature. Fe nanoclusters are formed on top of the surface superstructure creating a regular array with the period of the superstructure. We also demonstrate that at least the initial stage of Fe growth occurs in two-dimensional mode. In the areas of the surface where the strain pattern is not formed, random nucleation of Fe was observed.

## Introduction

Studies of the self-assembly of nanoclusters into ordered arrays are of great fundamental and technological importance. Nanostructures exhibit many novel physical and chemical properties and are essential for a fundamental understanding of condensed matter. Moreover, self-assembled nanostructures have enormous potential for technological applications. Increased constraints in lithography technologies have stimulated interest in radically alternative approaches to fabricating structures at the nanometer-scale to ensure a continuous progress towards down-sizing of electronics devices. In this respect, self-assembly is a highly promising avenue. For example, it is known that quantum dots can be formed on semiconductor substrates as a result of the strain induced during the heteroepitaxial growth of lattice-mismatched materials. By controlling the dimension and density of these quantum dot arrays, materials with novel optical and electronic properties can be engineered [1,2].

Self-assembly of magnetic nanostructures is also a field of great interest. For example, magnetic media with enhanced in-plane magnetoresistance have been constructed utilizing self-assembled nanogranular magnetic films [3,4]. The magneto-optical response of $In_{0.6}Ga_{0.4}As$ quantum dots has also been measured [5].

The growth of metal nanostructures on oxide substrates is an area of particular importance, since nanostructures grown on metal substrates are not suitable for many applications involving electron transport measurements. This area of research is still mostly unexplored. Only recently the wetting behaviour of palladium grown on a thin FeO film has been demonstrated [6] and regular arrays of metal nanoclusters were successfully formed on a nanopatterned alumina substrate [7].

In this letter we demonstrate that a well ordered array of nanoclusters can be grown on the patterned surface of $Fe_3O_4(111)$. Importantly, the observed nanoclusters were grown and analysed at room temperature.

**Experimental Results**

The experiments were performed in ultrahigh vacuum (UHV) at room temperature, using scanning tunneling microscopy (STM), low-energy electron diffraction (LEED) and Auger electron spectroscopy (AES). A synthetic single crystal of $Fe_3O_4(111)$ was used in these experiments. The details of the sample preparation are described elsewhere [8] but in brief, the preparation consisted in annealing the sample in a oxygen atmosphere of $10^{-6}$ Torr at a temperature of about 1000 K and then cooling it down at this oxygen pressure.

As reported in our recent publication [8], under oxidizing preparation conditions the (111) surface of magnetite reconstructs into a well-ordered superstructure with a periodicity of 42 Å and three-fold symmetry, which could be clearly identified by LEED. A STM image of this superstructure is shown in Fig.1. Three distinct areas, marked as areas I, II and III can be identified. The atomic symmetry of the top layer of all three areas is identical but they differ in their interatomic periodicity. Areas II and III have a 2.8 Å average periodicity, while the periodicity of area I is about 3.1 Å, as we reported in [8]. This surface corresponds to the oxygen termination of the bulk magnetite (111), which reconstructs under intrinsic stress to create a long-range order. It should be noted here that the topographical features (the corrugation and even the shape of the different areas) seen in Fig.1 are very sensitive to the scanning conditions (tunneling bias, current, tip properties) and originate from an electronic effect rather than real geometry of the superstructure. The image in Fig.1 can be interpreted as an image of the closed-packed oxygen lattice with the local variations in the interatomic spacing according to the surface strain pattern [8] with the area I under tensile and the area III under compressive stress.

The superstructure is stable at room temperature and is not significantly affected by adsorbates even after a few days in a UHV environment. These characteristics make it a very suitable template for the growth of self assembled nanostructures. We have evaporated an Fe film by electron beam evaporation. The evaporation was carried out at room temperature at a pressure of $3.7\times10^{-10}$ Torr. A Fe films 0.5 Å and 0.2 Å thicknesses were evaporated at a deposition rate of 0.5 Å/min. The film thickness and the deposition rate were measured by a quartz crystal balance. LEED measurements carried out after the deposition confirmed the presence of a 42 Å superlattice.

Fig. 2 shows a STM image of the Fe nanoclusters nucleated in ordered fashion on top of the superstructure, which is still clearly visible. The thickness of the nanoclusters was measured to be 2.2±0.3 Å. By using molecular simulation for the growth of Fe on the oxygen terminated $Fe_3O_4$(111) surface, we have established that Fe-films grow preferentially along the (110) Fe-bulk plane. The Fe (110) plane interatomic periodicity is 2.48 Å and the interlayer separation 2.03 Å. We therefore conclude that the Fe-islands represent one monolayer of iron. It is worth noting that the underlying superstructure is stable and is not affected by the deposition of the Fe film, and that the array of nanoclusters is stable at room temperature. It is clear from the STM image on Fig.2a that the Fe-islands nucleate preferentially on particular areas of the superstructure, creating an ordered array with the same periodicity as the superstructure. Further analysis of the STM images of the lower amount of deposited iron (Fig.2b) indicates that the nucleation is more likely to start on the areas III.

The demonstration that nucleation of the nanoclusters depends on surface structure is given by Fig. 3. Two terraces **A** and **B** separated by a step height of 2.5±0.4 Å correspond both to oxygen termination: one is reconstructed (**A**), and another contains a number of surface defects, which inhibit the superstructure formation (**B**) (more detailed data about magnetite (111) terminations can be found in [8]). The Fe nanoclusters form a regular array on terrace **A** only, while terrace **B** contains only randomly nucleated Fe islands. Terraces with no superstructure also exhibit a Fe decoration of the terrace edges, indicating that higher diffusion rate on those terraces allows Fe-adatom diffuse toward areas of higher surface energy.

**Discussion and Conclusions**

Previous studies suggest that oxide surfaces do not support 2*D* growth of metal films due to their low surface energy. Only recently, two-dimensional growth of Pd on a FeO (111) thin film surface was successfully demonstrated in [6]. Furthermore, wetting behaviour was theoretically predicted for Pd deposited on MgO(111) [9]. The similarities between the MgO(111), FeO(111) and $Fe_3O_4$(111) substrates lie in the fact that they are all type 3 polar surfaces with a close-packed oxygen layer on the top surface. Goniakovski and Noguera proposed that the 2*D* growth of Pd on MgO is due to a significant electron transfer between the deposited metal and the oxygen at the surface, which significantly enhances the adhesion energy [9]. A similar behaviour could be expected for the deposition of Fe on the oxygen terminated $Fe_3O_4$ (111) surface.

To the best of our knowledge, well defined metal nanostructures were successfully grown on reconstructed metal oxides in only two cases. In one case, Pd grown on an alumina film exhibits a self-ordering behavior, forming array of nanoclusters with a period equal to that of the surface superstructure [7]. In the second case, the FeO film used in Shaikhutdinov's et al. experiment as a substrate possesses a long range order, which is attributed to the lattice mismatch between the FeO film and Pt (111) substrate [6]. Although the authors did not claim that the growth takes place on preferential nucleation sites, the shape of the islands and the correlation between the superstructure periodicity and the island's size are a strong indication that this is the case. In both cases, the oxide thin film and substrate form a coincidence structure as a result of the lattice mismatch between them. Similarly to our case, one of the characteristics of such structures is a long-range modulation of the surface strain, which is always responsible for local changes in the diffusion barrier [10]. Although, a diffusion barrier increase on areas under tensile stress has been observed for the growth on metal substrates [11-13], which is contrary to our observation. However, the long-range surface strain modulation observed in our case leads to the difference in coordination between the top layer and the next layers (bulk) in areas I, and III. It is well known that the materials with different stacking sequence have different adsorption properties [14]. This could well be the reason for the formation of the preferential nucleation sites as opposed to a mere alteration of the diffusion barrier and binding energy by a uniform strain.

To summarize, the results presented here can be divided in two parts. First, the patterned surface of magnetite (111) presents preferential nucleation sites for the formation of Fe nanoclusters.

Second, as the reported Fe nanoclusters are one monolayer thick, one could conclude that at least the initial stages of 2D growth of a metal can be achieved on the magnetite (111) surface.

# References

1. G. Karczewski, S. Mackowski, M. Kutrowski, T. Wojtowicz, J. Kossut; Appl. Phys. Lett 74 (20) 3011-3013 (1999)
2. I. L. Aleiner, V.I. Fal'ko; Phys. Rev. Lett 87 (25) Art No 256801 (2001)
3. Y. Peng, H.L. Zhang, S.L. Pan, H.L. Li, Journal of Applied Physics 87, 7405-7408 (2000) ;
4. S. Mitani, S. Takahashi, K. Takanashi, K. Yakushiji, S. Maekawa, H. Fujimori, Phys. Rev. Lett. 81 (13) 2799-2808 (1998)
5. M.Bayer, A, Kuther, A. Forchel, A. Gorbunov, V.B. Timofeev, F. Schafer, J.P. Reitmaier, T.L. Reinecke, S.N. Walck, Phys. Rev. Lett. 82 (8) 1748-1751 (1999)
6. Sh.K. Shaikhutdinov, R.Meyer, D. Lahav, M, Baumer, T. Kluner, H.-J. Freud, Phys.Rev. Lett. **91**, 76102 (2003)
7. S. Degen, C. Becker, K. Wandelt, Faraday Discuss. 125, 343 (2004)
8. N. Berdunov, S. Murphy, G. Mariotto, and I. V. Shvets, Phys. Rev. B accepted, (2004); also arXiv:cond-mat/0403238
9. Goniakowski J, Noguera C, PRB **66**, 85417 (2002)
10. Larsson M.I., Sabiryanov R.F., Cho K., Clemens B.M., Surf.Sci.Lett. 536 (2003) L389
11. H. Brune, K. Bromann, H. Röder, K. Kern, J. Jacobsen, P. Stoltze, K. Jacobsen and J. Nørskov, Phys. Rev. B 52, 14380 (1995)
12. C. Ratsch , A.P. Seitsonen and M. Scheffler, Phys. Rev. B 55, 6750 (1997)
13. I.V. Shvets, S. Murphy et al., PRB submitted (2004); also arXiv:cond-mat/0405148
14. Brune H., Giovannini M., Bromann K., Kern K., Nature **394**, 451 (1998)

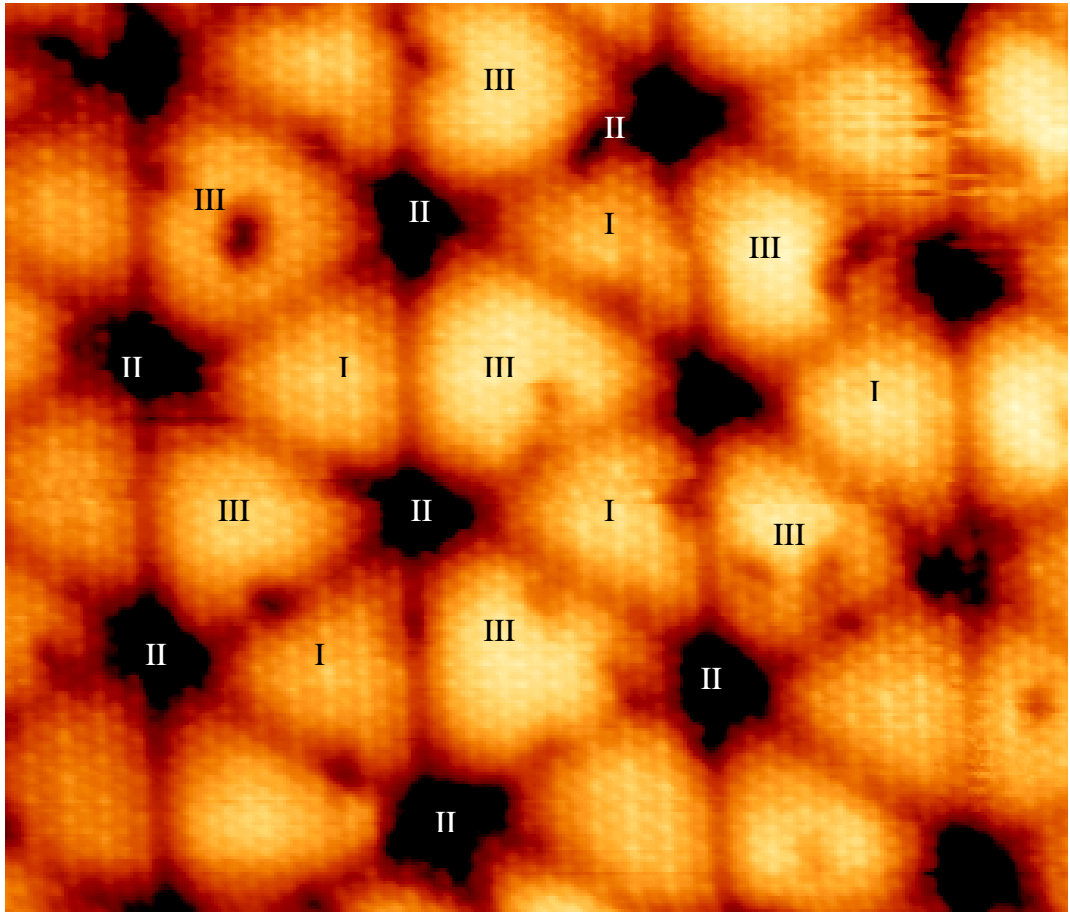

**Figure 1. (1500Å×1300Å) STM image of the surface superstructure. Three distinct areas are marked as I, II and III. The periodicity of the superstructure is 42 Å.**

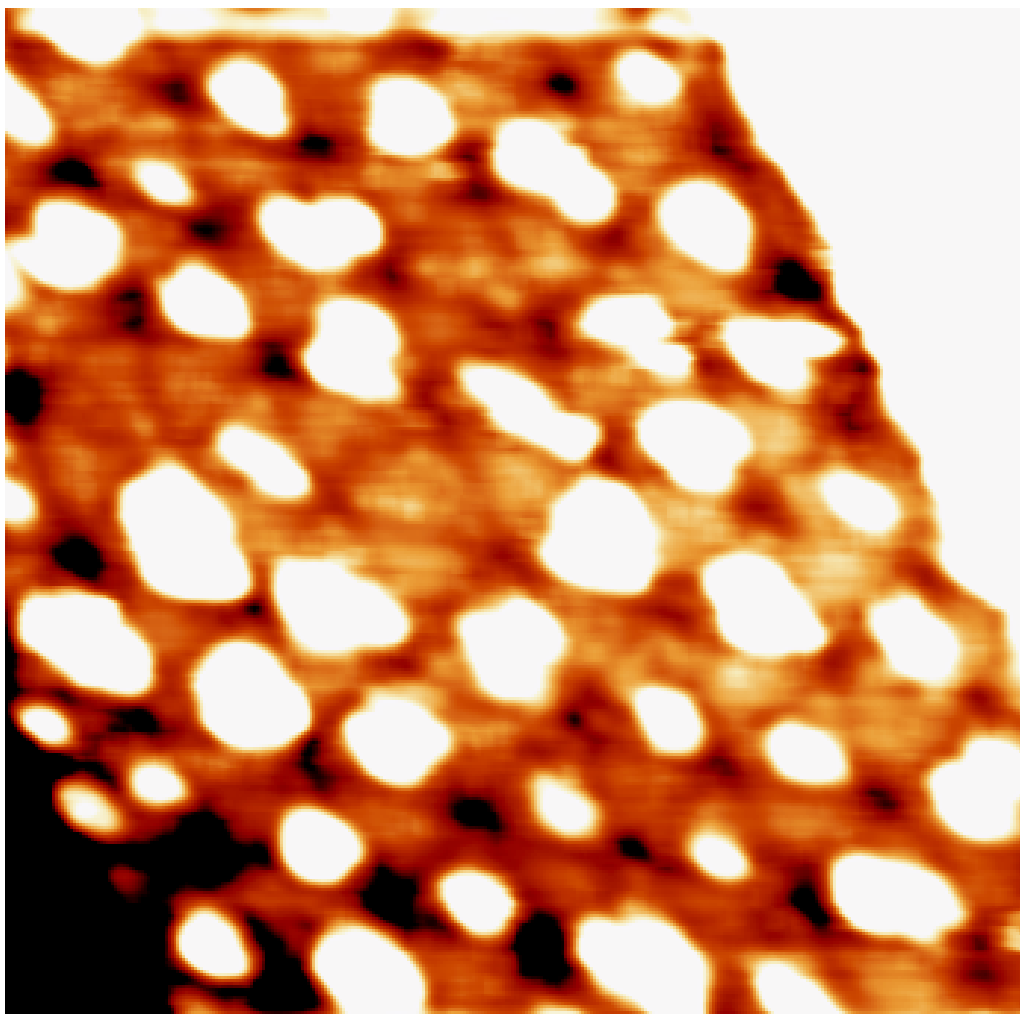

**Figure 2a. (300Å×300Å) STM image of an ordered array of Fe islands formed on the superstructured magnetite surface. Fe islands (white blobs) of 1 monolayer thickness are nucleated in the areas III of superstructure (Fig.1).**

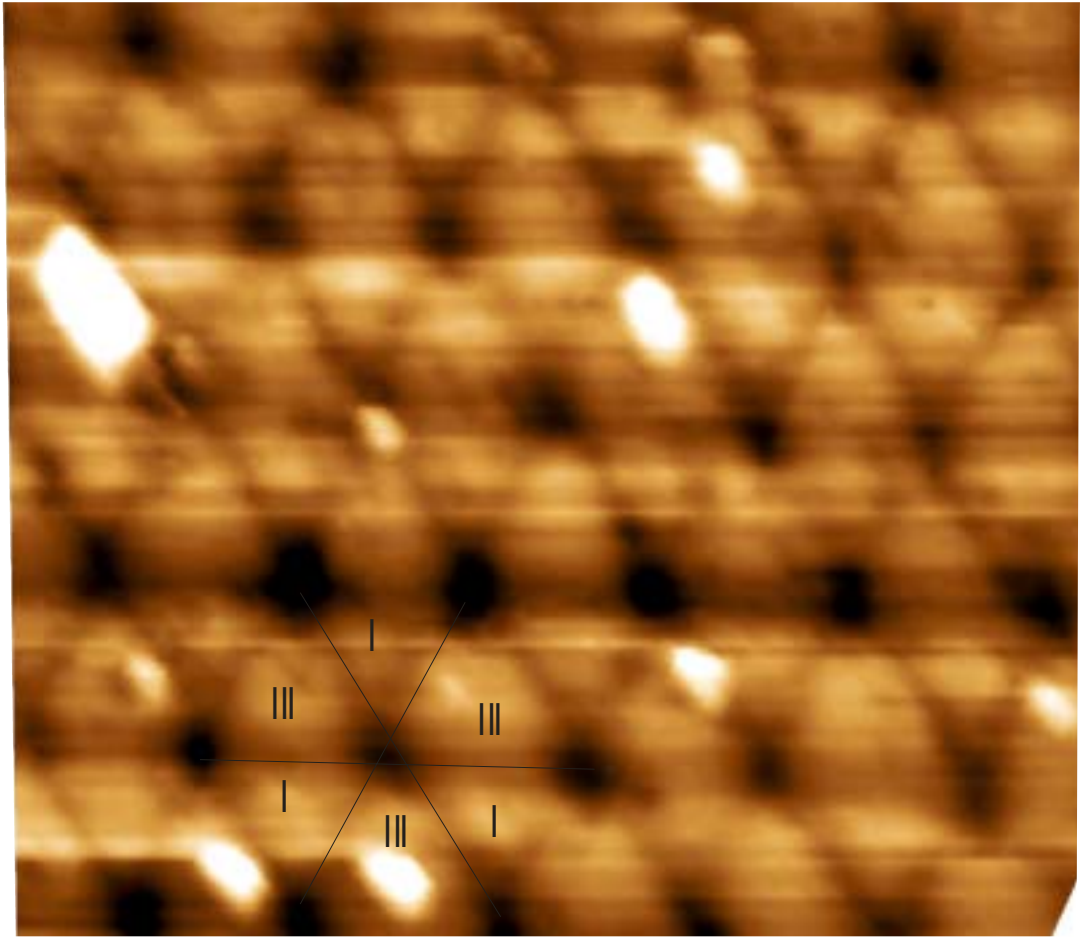

**Figure 2b. (250Å×220Å) STM image of 0.2 Å Fe film deposited on superstructured magnetite surface. Fe islands start to nucleate on the area III.**

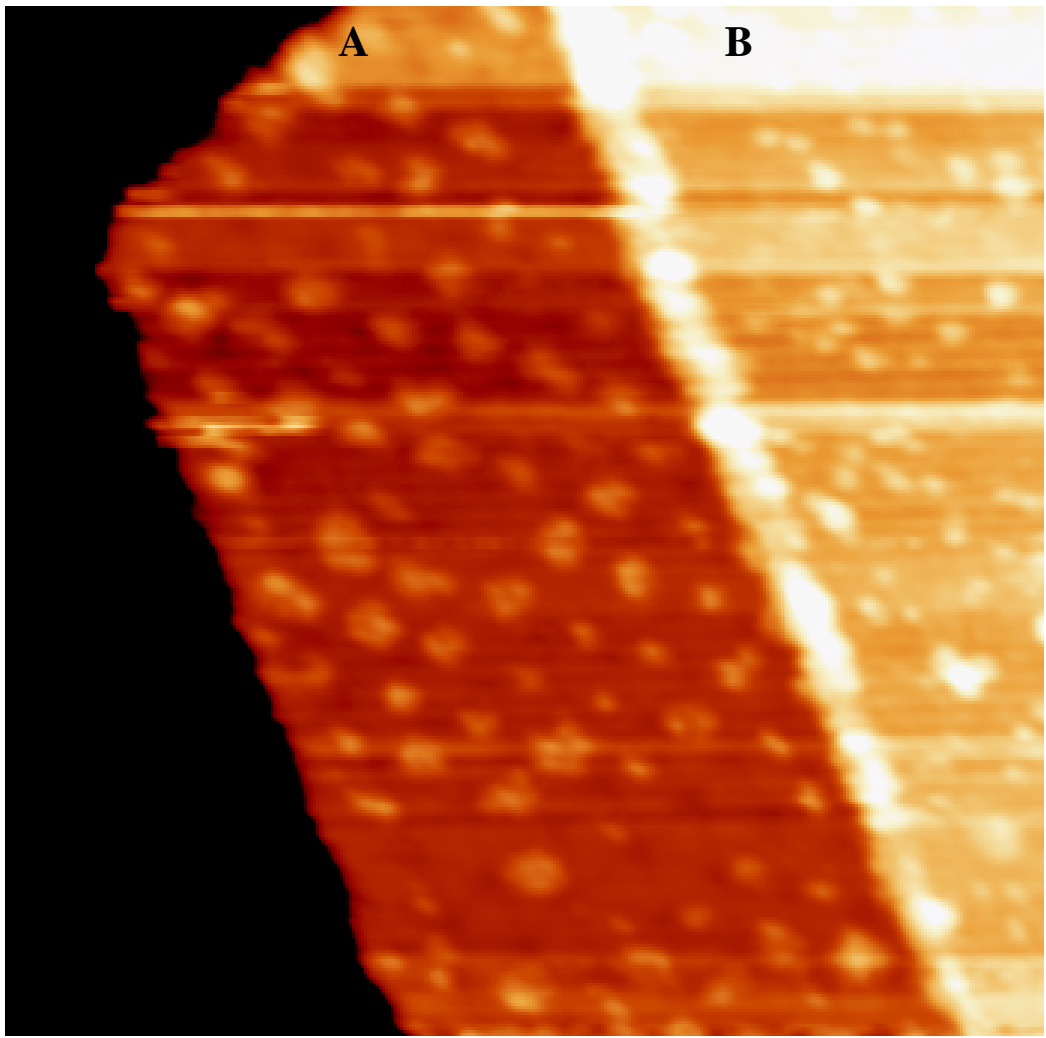

**Figure 3**. (600Å×600Å) STM image. Two visible terraces correspond both to oxygen termination: one is reconstructed to the superstructure (A), and another contains number of surface defects protecting superstructure formation (B). An ordered array of Fe islands has nucleated on top of the terrace A. The Fe islands have conserved the 42 Å periodicity of the underlying superstructure. The terrace B contains randomly nucleated Fe islands nucleated.